# FROM RELATIVISTIC ELECTRONS TO X-RAY PHASE CONTRAST IMAGING*

A. H. Lumpkin[1#], A.B. Garson[2], and M.A. Anastasio[2]
[1]Fermi National Accelerator Laboratory, Batavia, IL 60510 USA,
[2]Washington University in St. Louis**, St. Louis, MO 63130 USA

*Abstract*

We report the initial demonstrations of the use of single crystals in indirect x-ray imaging for x-ray phase contrast imaging at the Washington University in St. Louis Computational Bioimaging Laboratory (CBL). Based on single Gaussian peak fits to the x-ray images, we observed a four times smaller system point spread function (21 μm (FWHM)) with the 25-mm diameter single crystals than the reference polycrystalline phosphor's 80-μm value. Potential fiber-optic plate depth-of-focus aspects and 33-μm diameter carbon fiber imaging are also addressed.

## INTRODUCTION

X-ray phase contrast (XPC) imaging is an emerging technology that holds great promise for biomedical applications due to its ability to provide information about soft tissue structure [1]. The need for high spatial resolution at the boundaries of the tissues is noted for this process. Based on results on imaging of relativistic electron beams with single crystals [2], we proposed transferring single-crystal imaging technology to this bio-imaging issue. We report initial indirect x-ray imaging tests that demonstrated improved spatial resolution with single crystals compared to the $Gd_2O_2S$:Tb polycrystalline phosphor in a commercial, large-format CCD system. Using the Washington University microfocus x-ray tube as a source of 17 keV x-rays and the exchangeable phosphor feature of the camera system, we compared the point spread function (PSF) of the system with the reference phosphor to that with several rare-earth-garnet single crystals of varying thickness borrowed from the Fermilab and Argonne National Laboratory (ANL) linac labs.

We used a series of x-ray collimators which ranged in diameter from 400 to 25 microns. These were placed on the camera's Be entrance window to explore the PSF effects. Based on single Gaussian peak fits to the x-ray images, we observed a four times smaller system PSF (21 microns (FWHM)) with the 25-mm diameter single crystals than with the reference polycrystalline phosphor's 80-micron value. Initial images of 33-micron diameter carbon fibers have also been obtained with the system. The tests with a full-scale 88-mm diameter single crystal which would be fiber optically coupled to the CCD sensor with 86-mm diameter are being planned.

## TECHNICAL CONSIDERATIONS

The improved spatial resolution with single crystals over polycrystalline or powder samples had been previously noted in the imaging of relativistic electron beams [2]. Examples are shown in Fig. 1 as deduced from results at various laboratories, and the concept is being applied to indirect x-ray imaging in this research.

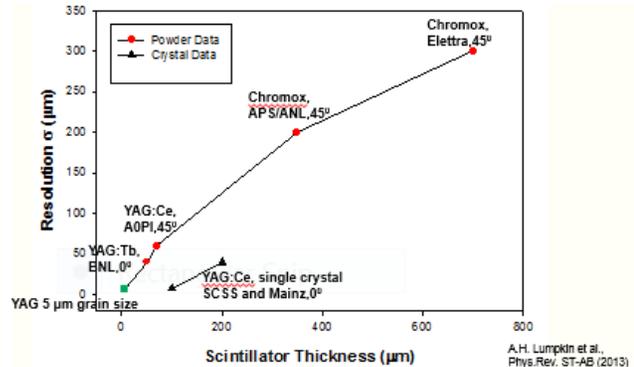

Figure 1: Summary of spatial-resolution values for different polycrystalline and single-crystal sample thicknesses for relativistic electron beams.

*The CBL facility*

The CBL XPC lab [3] includes the microfocus x-ray source, a high precision stage and rail system, and the high resolution Quad-RO x-ray camera as schematically shown in Fig. 2. The x-ray source is a Kevex PXS10-65W with cone beam, tungsten anode, 7-100 micron spot sizes, and 45-130 kV tube voltages. The Thorlabs rail system was used, but not the computer controlled stages. The Quad-RO-4096 is a Peltier cooled (-40 degrees C) CCD, with 15 micron pixel pitch for a 4096 x4096 array [4]. It has 14 bit intensity quantization and a PSF to be determined (generally 30-40 microns was ascribed).

We placed sequentially the collimators from the Amptec x-ray spectrometer set, on a lead plate with a hole drilled in it smaller than the W disc diameter. This plate was leveled with shims against the outer flange surface of the Quad-RO camera and positioned for the x-ray images to fall in the central area of one of the four quadrants of the CCD array. The set included collimators of 400, 200, 100, 50, and 25 μm in diameter. We estimated the rms size by dividing by the SQRT 12, and we multiplied by 2.35 to obtain the FWHM of the effective x-ray source size as tabulated in Table 1. The x-ray source was run at 7-μm spot size, at 25 or 60 kV tube voltage, at 150 μA current, 3.8 W, and located 0.9 m from the camera. Typical image integration times were 30s, and we averaged over five frames. Data were dark current subtracted, but only any mesh images were flat-field corrected. They were acquired and displayed with PI software [5]. Typically only a 400 x 400 region of interest (ROI) was then selected for processing in FNAL's image processing program, ImageTool, a MATLAB based program [6]. This program fits to Gaussian profiles the projected profiles from the selected smaller ROI and provides the amplitude, mean position, sigma, and the corresponding errors for each from the analyses. Background fit options are linear, flat, and quadratic. The program can fit up to 8 different peaks in the ROI, and we used this feature to assess the modulation of the wire grid and mesh data (not shown). It also provided the option for fitting projected profiles to a double Gaussian when that issue arose.

___________________
*Work partly supported under Contract No. DE-AC02-07CH11359 with the United States Department of Energy.
**Work at Washington University in St. Louis was supported in part by NSF CBET1263988.
#lumpkin@fnal.gov

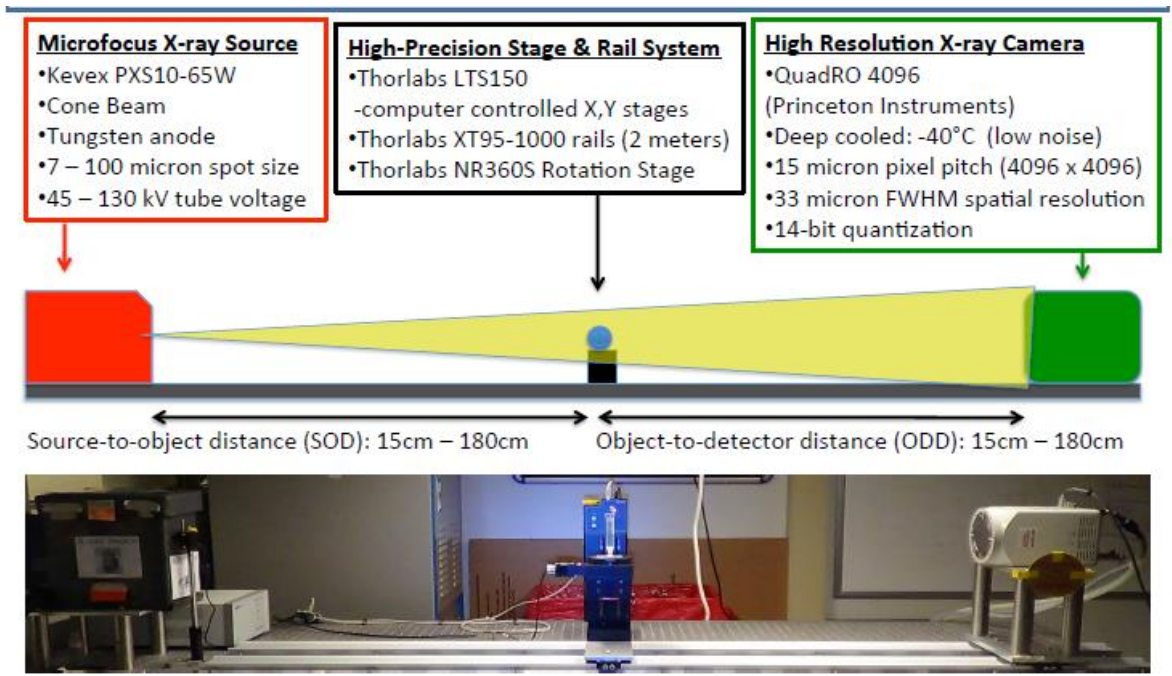

Figure 2: Schematic and photograph of the Washington University XPC imaging laboratory showing the microfocus x-ray source, the sample translation stages, and the QuadRO-4096 camera [3].

*The Single Crystals*

The rare-earth-garnet single crystals of 25-mm diameter were borrowed from the accelerator laboratories of FNAL and ANL. We obtained YAG:Ce, LUAg:Ce and LYSO:Ce crystals of thicknesses of 50, 100, and 200 μm. Two paired samples were obtained from Crytur, Inc. with Al coating as an optical reflector for the 50- and 100-μm examples. These were used to assess the role of the input FOP's depth of focus on the system PSF and signal gain.

Table 1: Summary of the x-ray collimators used to assess the PSF with 17-keV x-rays.

| Pinhole # | Diam (μm) | FWHM (μm) |
|---|---|---|
| 1 | 25 | 17 |
| 2 | 50 | 34 |
| 3 | 100 | 68 |
| 4 | 200 | 136 |
| 5 | 400 | 272 |

## X-RAY IMAGING RESULTS

*Collimated X-ray Image results*

We show the initial results of the 50-μm diameter collimator images as an example in Fig. 3. We obtained the reference Al–coated P43 phosphor data first and immediately noted that the projected vertical profile of 88 μm indicated the PSF was larger than the expected 40 μm. After taking the whole collimator set data with the P43, we installed the YAG;Ce and LuAG:Ce 50-μm thick crystals in the QuadRO positioned over two diagonal quadrants of the 4-quadrant sensor. Both the YAG;Ce and LuAG:Ce crystals had image sizes of about 36±1 μm, very close to the calculated 34-μm FWHM for this collimator. Figure 4 shows a summary plot comparing the polycrystalline and single crystal results. Using the smallest aperture of 17-μm FWHM (case 5), we deduced the system PSF (found by subtracting out the aperture size in quadrature) was about 21 μm with the single crystals, 4 times smaller than that with the reference P43 phosphor.

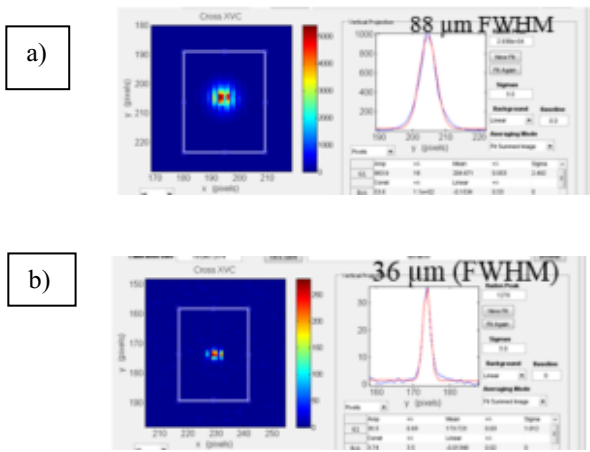

Figure 3: Initial images using the 50-µm diameter collimator with a) the reference P43 phosphor and b) a 50-µm thick single YAG:Ce crystal.

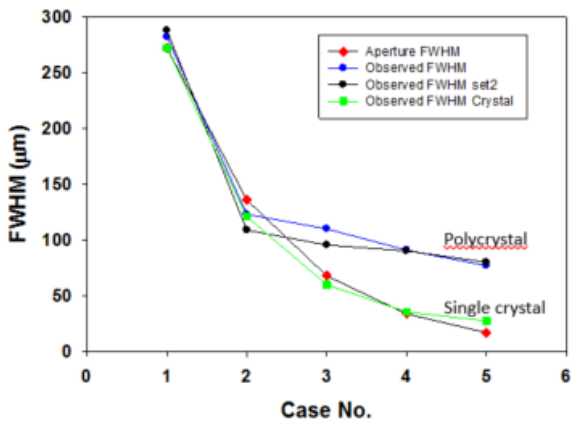

Figure 4: Plots of the measured projections for the different collimated images.

In a subsequent test series, we obtained two Al-coated YAG:Ce crystals, one with 50-µm thickness and one with 100-µm thickness. In Fig. 5 we show that the depth of focus of the input FOP plays a role in imaging with increasing effective optical thicknesses. The system PSF grows from 21 µm with 50-µm crystal thickness to about 70 µm with an effective optical thickness of 200 µm. The crystal PSF is much smaller than this system value [8]. So scintillator efficiency and this FOP term still need to be considered as a trade [7].

*Carbon Fiber PB-XPC Imaging Test*

As a simple test of the improved PSF with single crystals, we used propagation based (PB) XPC geometry to image the 33-µm diameter carbon fibers as shown in Fig. 6. The sharper image was obtained with the 50-µm thick crystal compared to that of the 100-µm thick crystal. We have since procured an 88-mm diameter YAG:Ce crystal bonded to a 90-mm FOP as shown in Fig. 7 [9]. This would be installed in the QuadRO-4096 camera for evaluation with a suitable phantom of bioimaging relevance.

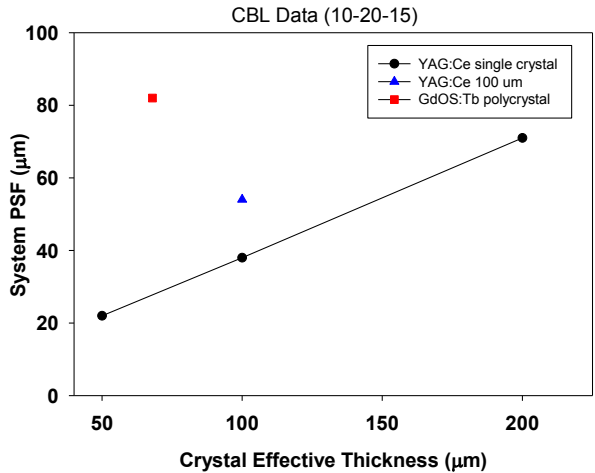

Figure 5: Plot of the system PSF vs. the crystal effective optical thickness showing the FOP depth-of-focus effect.

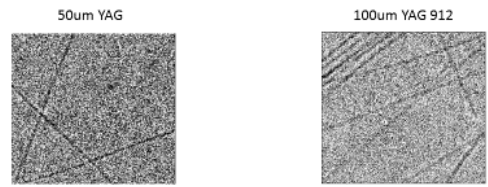

Figure 6: Propagation based XPC images of the 33-µm diameter carbon fibers with 50 µm (L) and 100 µm (R) thick single crystals.

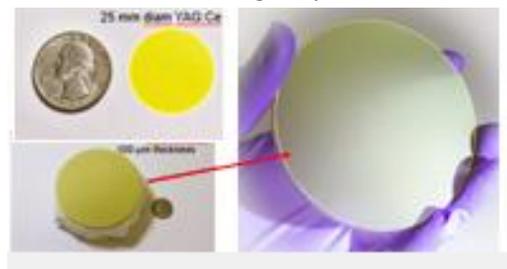

Figure 7: Photographs of the standard 25-mm diameter single crystal and the new 88-mm diameter crystal, *aka* "Katherine's Krystal" (photos by E. McCrory, FNAL).

## SUMMARY

In summary, we have performed initial studies of the improved spatial resolution obtained with single crystals compared to the reference P43 sample in the CBL XPC camera. We observed a 4 times better *system* PSF using the single crystals with the crystal PSF being even smaller. We have obtained an 88-mm diameter crystal bonded to a FOP to be integrated into this large format camera for final, full laboratory-scale PB-XPC tests.

## ACKNOWLEDGMENTS

The first author acknowledges the support of R. Dixon and N. Eddy at Fermilab.